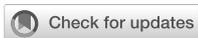

# Global visibility of publications through Digital Object Identifiers

Houcemeddine Turki[1], Grischa Fraumann[2]*, Mohamed Ali Hadj Taieb[1] and Mohamed Ben Aouicha[1]

[1]Data Engineering and Semantics Research Unit, Faculty of Sciences of Sfax, University of Sfax, Sfax, Tunisia, [2]R&D Department, TIB – Leibniz Information Centre for Science and Technology, Hannover, Germany

This brief research report analyzes the availability of Digital Object Identifiers (DOIs) worldwide, highlighting the dominance of large publishing houses and the need for unique persistent identifiers to increase the visibility of publications from developing countries. The study reveals that a considerable amount of publications from developing countries are excluded from the global flow of scientific information due to the absence of DOIs, emphasizing the need for alternative publishing models. The authors suggest that the availability of DOIs should receive more attention in scholarly communication and scientometrics, contributing to a necessary debate on DOIs relevant for librarians, publishers, and scientometricians.

## 1. Introduction

The availability of Digital Object Identifiers (DOIs) is of global relevance in publishing. Nevertheless, DOIs are not assigned to every publication, which limits the visibility of this subset in scholarly publishing. DOIs are a type of unique and global identifiers for digital objects, such as publications (Carter-Templeton et al., 2021). DOI registration agencies (e.g., *Crossref*) assign a DOI prefix to each publisher, which makes each article identifiable as an output of a specific publisher. Further parts of the DOI identify the venue (e.g., the journal) and the specific object (e.g., a journal article). While there are also other unique persistent identifiers, Digital Object Identifiers are important metadata elements in scholarly communication. But do all countries and their publications have DOIs? This is certainly not the case, as we will suggest below.

DOIs are used in scientometrics and related research fields, for example, to study the lists of references in publications (Mugnaini et al., 2021), or retrieve documents from repositories and match them with records in DOI registration agencies for citation analysis (Haupka et al., 2021). They can also be used to enrich bibliographic databases, such as *Scientific Electronic Library Online* (*SciELO*).[1] Additionally, DOIs can be used to conduct altmetric studies, that is, the perception of research outputs in online data sources, such as Wikipedia, Twitter, and more (Peters et al., 2016). However, DOIs are not allocated in certain journals and publishers in the Global South, except if researchers can publish their research in other international venues (e.g., journals and repositories) that provide DOIs.

---

1 https://images.webofknowledge.com/WOKRS513R8.1/help/SCIELO/hs_doi.html

To track the visibility and impact of scholarly publications, it is important to provide Digital Object Identifiers (DOIs) or other unique persistent identifiers for research outputs, particularly those issued by publishers in the Global South. In doing so, the visibility can be increased, for example, through a wider inclusion in altmetric sources and other sources that require unique persistent identifiers. This increased visibility was stressed in early work on altmetrics (Alperin, 2013). Do scientific publishers from the Global South have adequate DOI allocation? We want to raise awareness that the output of some scholarly publishers from the Global South is less visible in the "the global flow of scientific information" due to the lack of unique persistent identifiers, including DOIs (Mugnaini et al., 2021, p. 2524). This issue relates to previous work on the lower visibility of journals from the Global South, due to less inclusion in bibliographic databases, such as *Crossref* (Asubiaro and Onaolapo, 2023).

## 2. Availability of DOIs

We retrieved the list of DOI prefixes corresponding to journal publishers in *Crossref*.[2] It is true that there are DOI registration agencies beyond *Crossref*.[3] However, *Crossref* is among the largest ones, providing millions of DOIs (Hendricks et al., 2020) and having a significant representation of publishers from developing countries (Asubiaro and Onaolapo, 2023). This is why restricting our analysis to *Crossref* provides reliable results for our analysis. We decided to consider journal publishers instead of the institutions issuing conference proceedings and books/reference material reports by *Crossref*. We considered this publication type because journal articles are typically used as research data in scientometric studies. As of 17 January 2022, 98,420,414 DOIs and 103,606 source titles were reported by journal publishers. Some journal publishers and consequently DOI prefixes operate multiple journals. We only consider the 200 most published DOI prefixes including 83,472,052 DOIs (84.8%) and 37,833 scholarly journals (36.5%) for better computation and verification of data. This restriction will only have a minor influence on the output of our data collection and analysis as it captures the publishing behavior of most of the *Crossref* database. Our analysis is mainly based on the number of assigned DOIs and the considered DOI prefixes provide most of them. Afterwards, we used *OpenRefine*[4] to match metadata about the top 200 DOI prefixes in *Wikidata*,[5] an open and multidisciplinary knowledge graph providing large-scale bibliographic data (Nielsen et al., 2017), through the alignment of the publisher names with corresponding Wikidata items. A publisher can have more than one DOI prefix. But, this does not affect our analysis as we are interested in studying the whole picture of how DOIs are assigned and not in ranking the use of DOIs by different stakeholders.

When analyzing the top 200 DOI prefixes, we found out that the main DOI providers correspond to 15 large scholarly publishing houses, mostly created in the 19th century (See the *inception* column in the Table 1), such as *Elsevier* and *Springer* with a minor appearance of new publishing houses that publish open-access mega-journals such as *Public Library of Science* as shown in Table 1. This confirms the attraction of the scientific community to mega-journals due to their large research scope, rapid time to publication and their reach to a very broad audience (Björk, 2017). This also supports previous research findings about the domination of large publishing houses, particularly *Elsevier*, *Springer* and *Wiley*, on the market of scholarly publishing (Larivière et al., 2015). The oligopoly of scholarly journal publishing, which is mainly controlled by companies in developed countries, makes it difficult for developing countries to establish their own scholarly publishing traditions. This is because the publishing industry model is not adapted to the context of developing countries, which often lack funding, infrastructure, expertise, and research integrity (Posada and Chen, 2018).

The fact that developed countries are leading the scholarly publishing industry and research communities is verified by the following data. According to Figure 1 (gray bars), the United States of America, United Kingdom, Netherlands, Germany, Switzerland, France, and Japan are the main publishers of DOI items in *Crossref*. These countries are all located in the Global North, and they have a significantly higher representation in *Crossref* than domestic publishers in the Global South. This imbalance is due to a number of factors, including the long history of publishing houses in developed countries (Larivière et al., 2015; Posada and Chen, 2018), and the large market for scholarly publishing available in these countries (Posada and Chen, 2018). In recent years, however, there has been a growing trend of open access publishing in developing countries. This is motivated by a number of factors, including the increasing availability of funding for research, and the desire to increase the visibility of local research. As a result of this trend, some developing countries maintain several top 200 DOI prefixes, as depicted in Figure 1. Despite the efforts of several developing countries to expand their share in the scholarly publishing industry and assign DOIs to further publications, these nations failed to convince the worldwide research community to significantly contribute to their scholarly venues. This proves that developing countries face significant challenges to grow their scholarly publishing industries and this is what explains the gap between publishing houses in the main developed countries and the ones from the Global South (Salager-Meyer, 2008).

The disparities are not only restricted to the country representation of institutions issuing DOIs but also concerns the types of institutions providing DOIs. As shown in Table 2, scholarly publishers, scientific societies and non-profit organizations are the main establishments involved in assigning DOIs. University presses, research institutions, libraries, governments, and universities account for less DOIs in the present dataset, although some of them also provide their own scholarly publishing outlets. Asubiaro and Onaolapo (2023) also showed the relatively low share of university publishers of journals from developing countries in *Crossref*, compared to other categories of publishers. This occurs for a few reasons. First, these institutions typically publish large-scale reports, books, and book chapters (Ganu, 1999), which are more challenging to publish and disseminate than scholarly journals and conferences (Ali et al., 2013). Second, there are open-access DOI providers, such as data and publication repositories (e.g., *Zenodo*)

---

2 https://www.crossref.org/06members/51depositor.html
3 https://www.doi.org/the-community/existing-registration-agencies/
4 https://openrefine.org/
5 https://www.wikidata.org/

TABLE 1 Top 16 most published DOI prefixes in *Crossref* as of 17 January 2022.

| DOI prefix | Name | Instance of | Country | Journal count (percentage) | Total DOIs (percentage) | Inception |
|---|---|---|---|---|---|---|
| 10.1016 | Elsevier BV | Publisher | Netherlands | 4,262 (4.1%) | 17,218,689 (17.4%) | 1,880 |
| 10.1007 | Springer Science+Business Media | Publisher | Germany | 3,323 (3.2%) | 6,551,598 (6.6%) | 1,842 |
| 10.1002 | John Wiley & Sons Ltd | Publisher | United Kingdom | 1,358 (1.3%) | 5,380,888 (5.5%) | 1,807 |
| 10.1080 | Taylor & Francis | Publisher | United Kingdom | 3,736 (3.6%) | 4,392,461 (4.5%) | 1,852 |
| 10.1111 | Wiley-Blackwell | Publisher | United States of America | 1,382 (1.3%) | 3,642,267 (3.7%) | 2,001 |
| 10.1371 | Public Library of Science | Website | United States of America | 10 (<0.4%) | 3,478,859 (3.5%) | 2,000 |
| 10.1093 | Oxford University Press | University press | United Kingdom | 563 (0.5%) | 3,140,580 (3.2%) | 1,586 |
| 10.1177 | SAGE Publications | Book publisher | United States of America | 1,555 (1.5%) | 2,609,787 (2.7%) | 1,965 |
| 10.1021 | American Chemical Society | Scientific society | United States of America | 93 (<0.4%) | 2,163,704 (2.2%) | 1,876 |
| 10.1097 | Wolters Kluwer | Book publisher | Netherlands | 396 (<0.4%) | 1,893,239 (1.9%) | 1,987 |
| 10.1017 | Cambridge University Press | University press | United Kingdom | 613 (0.6%) | 1,633,902 (1.7%) | 1,534 |
| 10.2307 | JSTOR | Organization | United States of America | 748 (0.7%) | 1,603,832 (1.6%) | 1,995 |
| 10.1038 | Springer Science+Business Media | Publisher | Germany | 214 (<0.4%) | 1,364,997 (1.4%) | 1,842 |
| 10.1109 | Institute of Electrical and Electronics Engineers | Standards organization | United States of America | 397 (<0.4%) | 1,294,983 (1.3%) | 1,963 |
| 10.1136 | BMJ | Publisher | United Kingdom | 81 (<0.4%) | 923,126 (0.9%) | 1,840 |
| 10.1088 | IOP publishing | Publisher | United Kingdom | 121 (<0.4%) | 914,137 (0.9%) | 1,874 |

Springer Science+Business Media has two separate DOI prefixes in this sample. Wiley-Blackwell is a business of John Wiley & Sons Ltd.

that do not charge a fee for DOI allocation. In contrast, direct registration of DOIs in *Crossref*, the main DOI provider, is subject to a fee even for non-profit organizations and public institutions.[6] This can be a barrier for these institutions, which may struggle with funding and online payment of fees. DOIs are generated, for instance, by registering a metadata record at *Crossref*.[7] This registration process is only available for *Crossref* members, but does not differ based on geographical location of the publisher. This structure of fees might be different in other contexts that we did not consider in this brief research output with a focus on *Crossref*. Further limitations of the present study include that the overall numbers of assigned DOIs per country are not compared to the overall numbers of research outputs per country. The number of research outputs per country is also related to the number of researchers per country, which can vary to a high degree across countries. Furthermore, the location of publishers as shown in Figure 1 does not necessarily reflect the affiliation of authors. Such comparisons would be valuable, but are out of scope of this brief research report.

## 3. Conclusion

In conclusion, Digital Object Identifiers (DOIs) play a critical role in the accessibility and discoverability of online publications, but their availability is not equally distributed across the world. Our analysis of the top 200 DOI prefixes registered with *Crossref* reveals a dominance of large publishing houses from high-income countries in North America and Europe, with limited representation from the Global South. This has significant implications for global scholarly communication, including the visibility and adoption of metrics and indicators, and the need for alternative solutions and infrastructures. Therefore, we urge the scholarly community to address these issues by promoting the availability of DOIs globally and fostering a more inclusive and equitable scholarly communication system. Initiatives that try to tackle these issues, such as the Global Equitable Membership (GEM) program launched by *Crossref*[8] after the data collection of the present study, point toward the right direction and can make publications from several countries of the Global South, among others, more visible. Similarly, we would like to encourage more representatives from the Global South to join the DOI Foundation,[9] which would help to raise the visibility of research originating from a large part of the world. While this membership is not a requirement to allocate DOIs for publications, it would support the development of the global scholarly publishing system. Finally, planned DOI registration agencies, such as those by the Africa Persistent Identifier (PID) Alliance (Ksibi et al., 2023), that are tailored to the publications of specific world regions can increase the visibility of publications globally. This could be a crucial step to assign more DOIs to publications from the Global South.

---

6 Further information can be found at https://www.crossref.org/fees/.

7 Further information can be found at https://www.crossref.org/services/content-registration/.

8 Further information can be found at: https://www.crossref.org/gem/.

9 Further information can be found at: https://www.doi.org/the-community/who-are-the-members-and-users/.

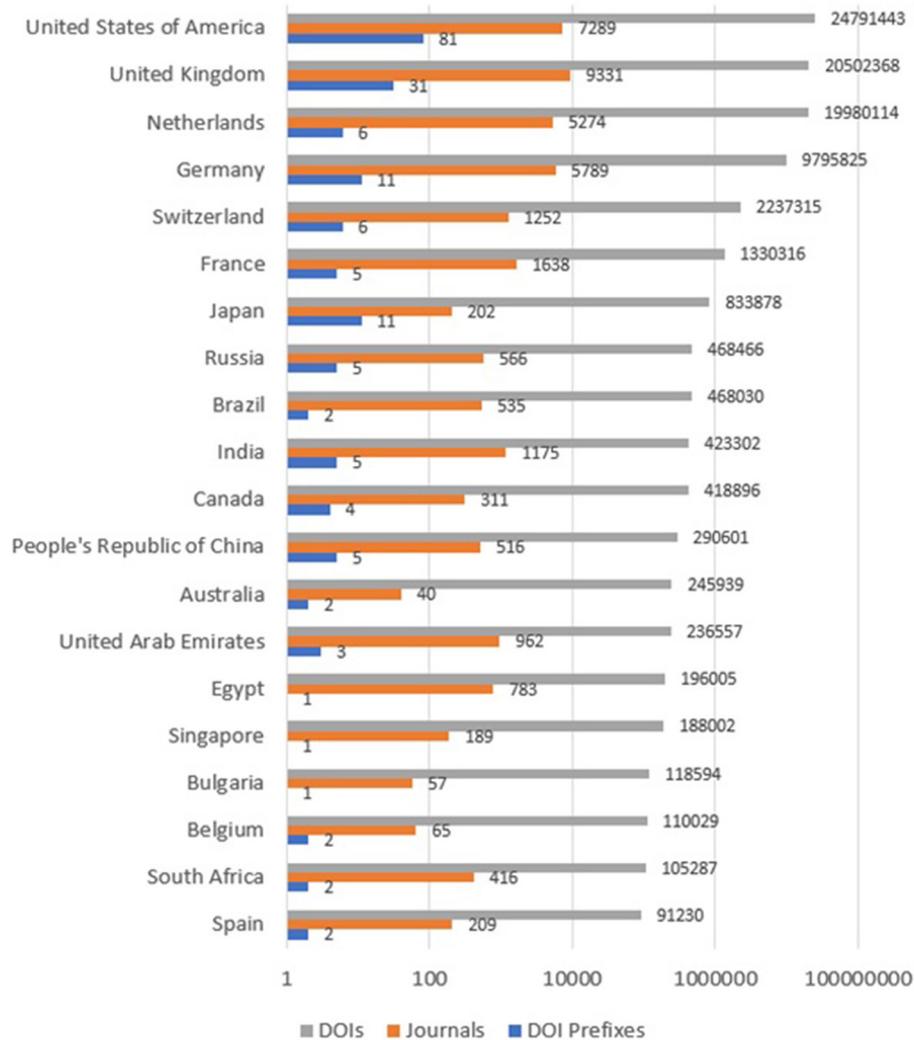

FIGURE 1
Top 20 countries assigning DOIs based on the *Crossref* Top 200 DOI Prefixes as of 17 January 2022.

TABLE 2 Types of institutions issuing *Crossref* DOIs (200 top DOI prefixes) as of January 17, 2022.

| Type | DOI prefixes | Journals | DOIs |
|---|---|---|---|
| Publisher | 84 | 28,696 | 55,441,839 |
| Scientific society | 54 | 1,532 | 9,192,417 |
| Organization | 29 | 1,613 | 6,281,437 |
| University press | 8 | 1,645 | 5,641,068 |
| Repository | 8 | 2,425 | 5,562,784 |
| Journal series | 7 | 22 | 644,533 |
| Research institutions and libraries | 8 | 1,022 | 457,894 |
| Government | 2 | 878 | 250,080 |

# Data availability statement

The datasets presented in this manuscript have been uploaded to the GitHub repository and can be accessed via the following link: https://github.com/csisc/DOIPrefixAnalysis.

# Author contributions

All authors contributed equally to this manuscript in its conception, writing, and editing.

# Funding

The publication of this article was funded by the Open Access Fund of Technische Informationsbibliothek (TIB) and the research was funded by the German Federal Ministry of Education and Research (BMBF) under grant number 01PU17019.

# Acknowledgments

We thank the peer reviewers for their valuable comments that improved the manuscript. We also thank the speakers and attendees of the 2019 Latin American Symposium on the Metrics Studies of Science and Technology (LASMSST) in Mexico City and the 2019 ENRESSH (European Network for Research Evaluation in the Social Sciences and Humanities) training school in Poznań for discussing the availability of DOIs. We also thank Stephanie Hagemann-Wilholt (TIB – Leibniz Information Centre for Science and Technology) and Zohreh Zahedi (CWTS, Leiden University) for commenting on an earlier draft of this article.

# Conflict of interest

The authors declare that the research was conducted in the absence of any commercial or financial relationships that could be construed as a potential conflict of interest.

The handling editor RM declared a past co-authorship with the author GF.

# Publisher's note

All claims expressed in this article are solely those of the authors and do not necessarily represent those of their affiliated organizations, or those of the publisher, the editors and the reviewers. Any product that may be evaluated in this article, or claim that may be made by its manufacturer, is not guaranteed or endorsed by the publisher.